\documentclass{webofc}

\usepackage[varg]{txfonts}   
\usepackage{hyperref}
\usepackage{url}
\usepackage{lineno}
\usepackage[Symbol]{upgreek}
\newcommand{\bbbar}{\ensuremath{\mathrm{b\overline{b}}}\xspace}
\newcommand{\bz}{\ensuremath{\mathrm{B^{0}}}\xspace}

\newcommand{\bp}{\ensuremath{\mathrm{B^{+}}}\xspace}
\newcommand{\dm}{\ensuremath{\mathrm{D^{-}}}\xspace}
\newcommand{\dz}{\ensuremath{\mathrm{D^{0}}}\xspace}

\newcommand{\pt}{\ensuremath{p_{\mathrm{T}}}\xspace}

\newcommand{\tev}{\ensuremath{\mathrm{TeV}}\xspace}
\newcommand{\gevc}{\ensuremath{\mathrm{GeV}/c}\xspace}

\newcommand{\pb}{\ensuremath{\mathrm{pb}}\xspace}

\newcommand{\sqrts}{\ensuremath{\sqrt{s}}\xspace}
\newcommand{\ppbar}{\ensuremath{\mathrm{p\overline{p}}}\xspace}

\hypersetup{colorlinks=true,citecolor=blue,urlcolor=blue,linkcolor=blue}

\begin{document}

\title{Testing perturbative QCD calculations with beauty-meson production in proton--proton collisions with ALICE}

\author{\firstname{Fabrizio} \lastname{Chinu}\inst{1,2}\fnsep\thanks{\email{fabrizio.chinu@cern.ch}} on behalf of the ALICE Collaboration}

\institute{Università degli Studi di Torino \and Istituto Nazionale di Fisica Nucleare}

\abstract{Measurements of the production cross section of beauty hadrons in proton--proton (pp) collisions provide excellent tests of perturbative quantum chromodynamics (pQCD) calculations. 
Theoretical approaches based on the factorisation theorem allow for the calculation of differential cross sections for hadron production as functions of transverse momentum (\pt) and rapidity ($y$). Measurements down to low transverse momenta are also fundamental ingredients for the estimation of the \bbbar production cross section. 

In this contribution, the measurement of \bz-meson production in pp collisions at $\sqrts=13.6$~TeV collected by the ALICE experiment during LHC Run 3 is presented. The \bz mesons are fully reconstructed via their decay channels into a D meson and a charged pion. The measured production cross section is compared with state-of-the-art pQCD calculations with next-to-leading order accuracy plus all-order resummation of next-to-leading logarithms.
}
\maketitle

\section{Introduction}\label{sec:introduction}
Because of their large mass, heavy quarks (i.e., charm and beauty) are exclusively produced in processes with large transferred momentum, and hence their production can be computed with pQCD calculations. The measurement of the production of heavy-flavour hadrons in pp collisions is therefore an excellent test of pQCD calculations. Calculations at the Next-to-Leading Order plus Next-to-Leading Log approximation successfully describe the measurements of both beauty hadron production~\cite{LHCb:2017vec, CMS:2024vip} and of non-prompt charm hadrons~\cite{ALICE:2024xln} (i.e. those produced from the decay of a beauty hadron) in pp collisions at the LHC. While open beauty production in pp collisions at LHC energies has been measured in a wide \pt range by the LHCb Collaboration~\cite{LHCb:2017vec} at forward rapidity, and at high \pt ($\pt>5$~\gevc) by the CMS Collaboration~\cite{CMS:2024vip} at midrapidity, the production of open beauty hadrons down to low \pt at midrapidity was never directly measured.

\section{Measurement of the production cross section of \texorpdfstring{$\mathrm{\mathbf{B^0}}$}{B0} mesons in pp collisions at \texorpdfstring{$\mathbf{\sqrt{s}=13}$~TeV}{sqrt(s)=13 TeV}}\label{sec:cross_section}
Exploiting the large dataset collected during the LHC Run~3 data-taking period, the ALICE Collaboration was able to perform the first measurement of the \bz production cross section at midrapidity ($|y|<0.5$) down to $\pt=1~\mathrm{GeV}/c$ at LHC energies~\cite{ALICE-PUBLIC-2025-004}. 
\begin{figure}[tb]
    \centering
    \vspace*{-1cm}
    \includegraphics[width=9.8cm,clip]{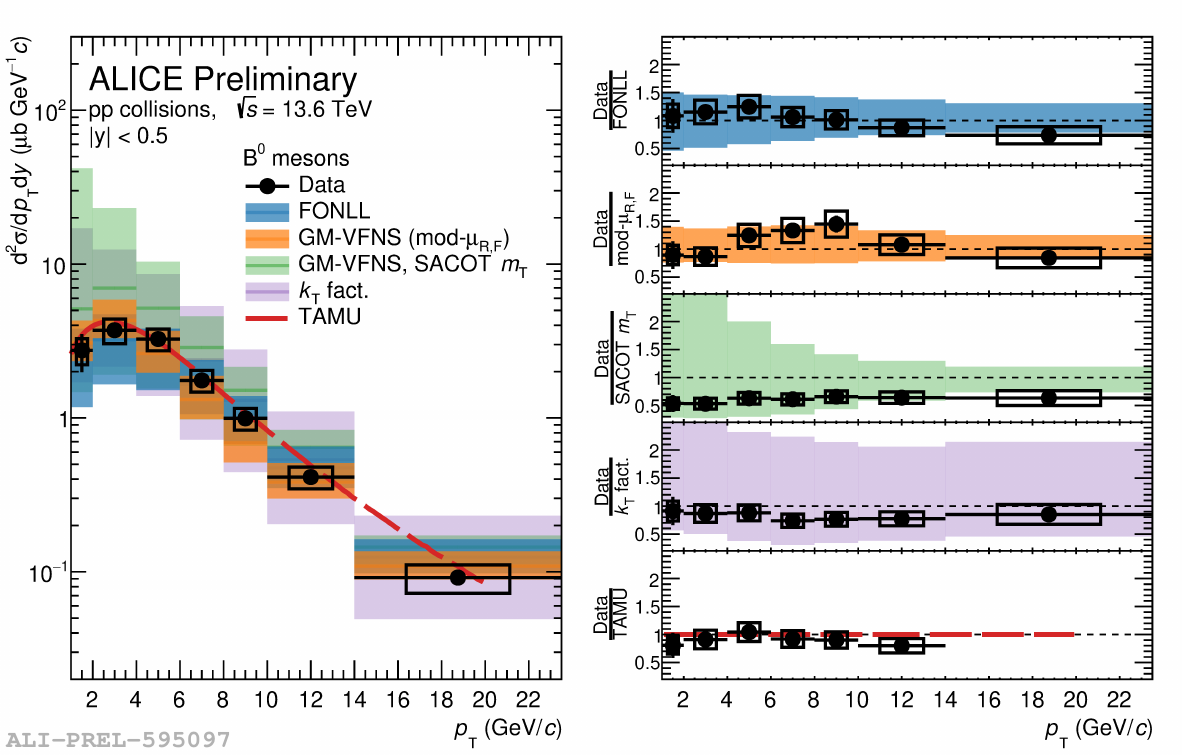}
    \caption{\pt-differential production cross section of \bz mesons measured in pp collisions at \mbox{$\sqrts=13.6$}~TeV compared with FONLL~\cite{Cacciari:2012ny}, GM-VFNS(mod-$\mu_\mathrm{R,F}$)~\cite{Benzke:2019usl}, GM-VFNS(SACOT-$m_\mathrm{T}$)~\cite{Helenius:2023wkn}, $k_\mathrm{T}$-factorisation~\cite{Barattini:2025wbo}, and TAMU~\cite{He:2022tod} calculations (left panel) and ratios of the data to the theoretical predictions (right panel).}
    \label{fig:xsec_pqcd}
    \vspace*{-1cm}
\end{figure}

The production cross section of \bz mesons is measured in the rapidity range $|y|<0.5$ and the $1 < \pt < 23.5$~\gevc interval, and is computed from the \bz raw yields, which include particles and antiparticles, extracted in each \pt interval via a fit to the invariant mass distribution of selected candidates. They are corrected for the detector acceptance and efficiency, branching ratio $\mathrm{BR} = \mathrm{BR (\bz\rightarrow D^-\pi^+)}\times\mathrm{BR (\dm\rightarrow \pi^-K^+\pi^-)} = (2.35 \pm 0.08)\cdot10^{-6}$~\cite{ParticleDataGroup:2024cfk} and integrated luminosity $\mathcal{L}_\mathrm{int} = (43 \pm 4) ~\pb^{-1}$. The cross section is shown in Fig.~\ref{fig:xsec_pqcd} as a function of \pt, and shows a good agreement with state-of-the-art pQCD calculations~\cite{Cacciari:2012ny, Benzke:2019usl, Helenius:2023wkn, Barattini:2025wbo}, as well as with predictions from the TAMU~\cite{He:2022tod} model, where the beauty-quark \pt spectrum is taken from FONLL calculations~\cite{Cacciari:2012ny} and a statistical approach for the description of the hadronisation process is used. The measured \pt-differential production cross section is found to be consistent with all the considered theoretical predictions within uncertainties.

\begin{figure}[ht]
    \centering
    \vspace*{-1cm}
    \includegraphics[width=10cm,clip]{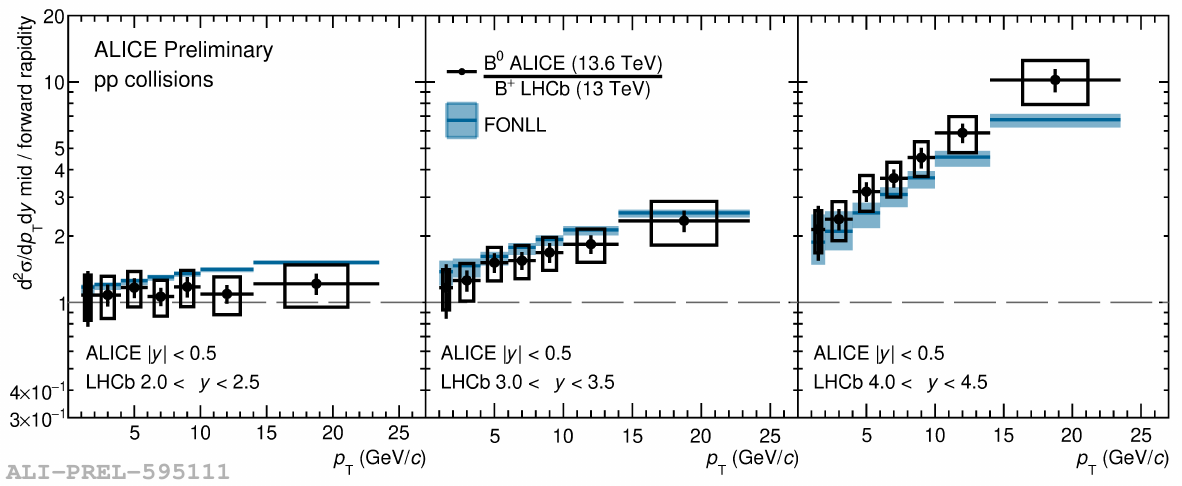}
    \includegraphics[width=10cm,clip]{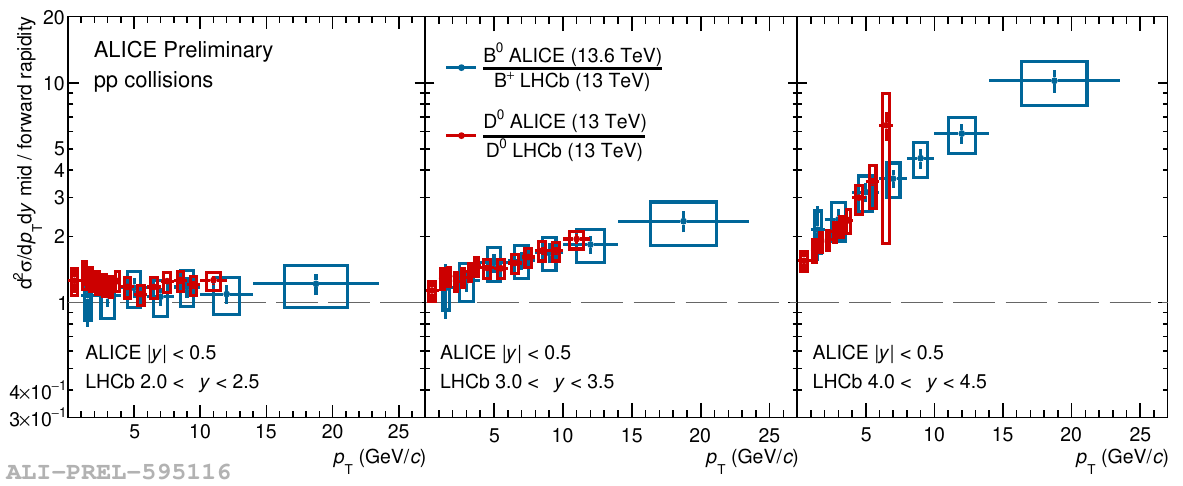}
    \caption{Ratios of \pt-differential production cross sections per unit of rapidity of \bz mesons at midrapidity in pp collisions at $\sqrts=13.6~\tev$ to those measured by the LHCb Collaboration~\cite{LHCb:2017vec} for \bp mesons in pp collisions at $\sqrts=13~\tev$ in three intervals of rapidity. The data points are compared to FONLL calculations~\cite{Cacciari:2012ny} in the first row and to the same ratio measured for \dz mesons in pp collisions at $\sqrts=13~\tev$~\cite{ALICE:2023sgl} in the second row.}
    \label{fig:mid_fwd_ratio}
    \vspace*{-0.5cm} 
\end{figure}

The rapidity dependence of the \bz meson production cross section is also studied by considering the ratio between the \bz meson production cross section measured at midrapidity by the ALICE Collaboration and the one measured for the \bp meson at forward rapidity by the LHCb Collaboration~\cite{LHCb:2017vec}. The ratio is shown in Fig.~\ref{fig:mid_fwd_ratio} and is found to follow different trends with \pt in the different rapidity intervals. It is well described by FONLL calculations~\cite{Cacciari:2012ny}. The ratio is also compared to the one measured for \dz mesons in pp collisions at $\sqrts=13$~TeV~\cite{ALICE:2023sgl}, which shows a similar trend with $y$, suggesting the factorisation of the mass and rapidity dependence of the production cross section of heavy flavour hadrons.

\begin{figure}
    \centering
    \sidecaption
    \vspace*{-0.2cm}
    \includegraphics[width=7.cm,clip]{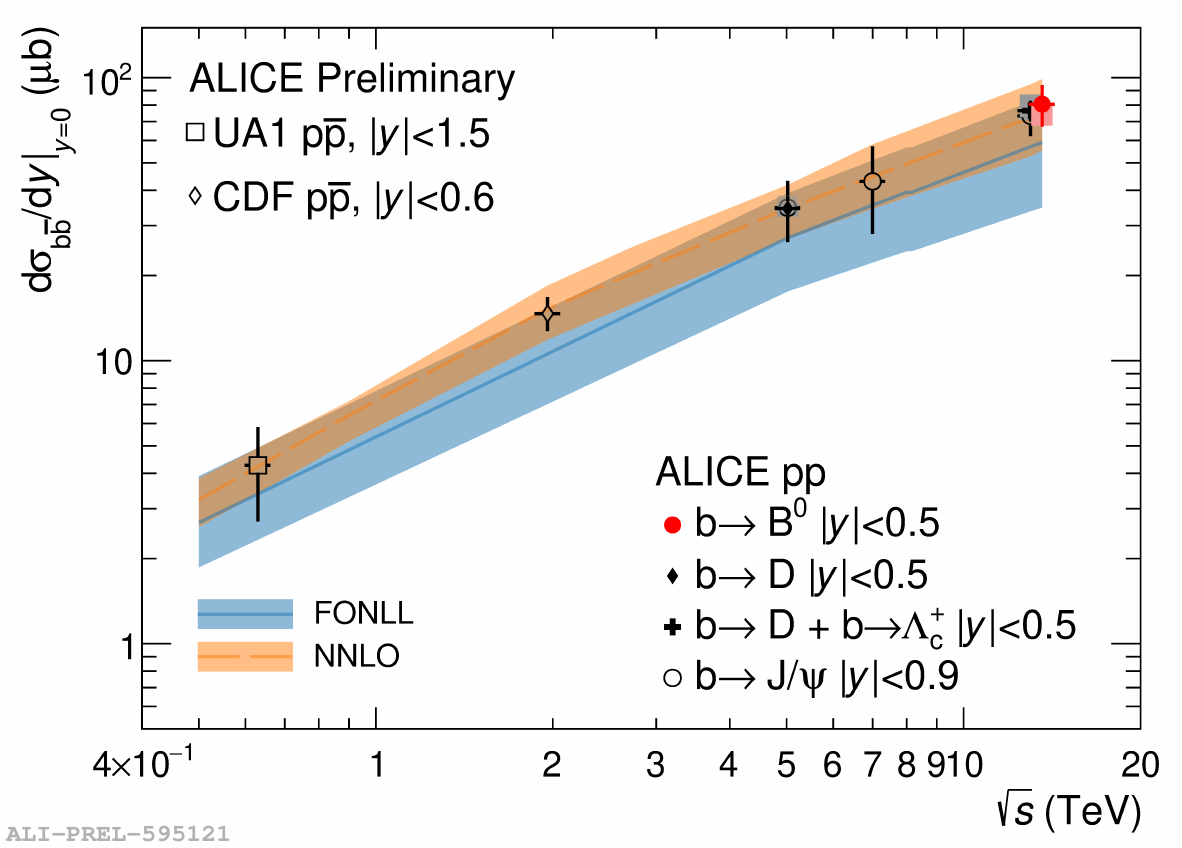}
    \caption{Beauty production cross section per unit of rapidity at midrapidity as a function of $\sqrts$ as measured in pp collisions by the ALICE Collaboration~\cite{ALICE:2021edd,ALICE:2021mgk,ALICE:2024xln,ALICE:2023wbx} and in \ppbar collisions by the CDF~\cite{CDF:2004jtw} and UA1~\cite{UA1:1990vvp} Collaborations. The solid lines with the shaded band represents the FONLL~\cite{Cacciari:2012ny} and NNLO~\cite{Catani:2020kkl} calculations with their uncertainties, respectively.}
    \label{fig:bbbar_xsec}
    \vspace*{-0.2cm} 
\end{figure}

The \bbbar production cross section at midrapidity ($|y|<0.5$) can be estimated starting from the total \bz production cross section and the fragmentation fractions measured in $\mathrm{p\overline{p}}$ collisions~\cite{HFLAV:2019otj}. The \bz production cross section can in turn be obtained by integrating the \pt-differential production cross section measured in the $1 < \pt < 23.5$~\gevc range and by applying an extrapolation factor obtained from FONLL predictions~\cite{Cacciari:2012ny}. The resulting \bbbar production cross section per unit of rapidity in pp collisions at $\sqrts=13.6$~TeV and $|y|<0.5$  is shown in Fig.~\ref{fig:bbbar_xsec}, and is found to be in line with the increasing trend with energy of previous measurements at lower energies. Predictions at NNLO precision~\cite{Catani:2020kkl} are available for the \bbbar production cross section, and are consistent with the measured value. They are found to be in good agreement with FONLL predictions~\cite{Cacciari:2012ny}, while laying on the upper edge of its uncertainty band.

\section{Conclusions}\label{sec:conclusions}
The first measurement of the production cross section of \bz mesons in pp collisions performed by the ALICE Collaboration was presented. The considered observables are found to be in good agreement with state-of-the-art pQCD-based calculations. This measurement provides a solid baseline for the study of beauty-hadron production in Pb--Pb collisions, where modifications due to the presence of the quark--gluon plasma are expected. Measurements of other beauty hadrons will allow the study of the relative abundances of different beauty-hadron species, providing insights about the beauty-quark hadronisation mechanisms.

\bibliography{bibliography} 

\begin{thebibliography}{18}

\bibitem{LHCb:2017vec}
R.~Aaij et~al. (LHCb), {Measurement of the $B^{\pm}$ production cross-section
  in pp collisions at $\sqrt{s} =$ 7 and 13 TeV}, JHEP \textbf{12}, 026 (2017),
  \texttt{1710.04921}. \doiwoc{10.1007/JHEP12(2017)026}

\bibitem{CMS:2024vip}
A.~Hayrapetyan et~al. (CMS), {Bottom quark energy loss and hadronization with
  B$^{+}$ and $ {\textrm{B}}_{\textrm{s}}^0 $ nuclear modification factors
  using pp and PbPb collisions at $ \sqrt{s_{\textrm{NN}}} $ = 5.02 TeV}, JHEP
  \textbf{02}, 195 (2025), \texttt{2409.07258}.
  \doiwoc{10.1007/JHEP02(2025)195}

\bibitem{ALICE:2024xln}
S.~Acharya et~al. (ALICE), {Measurement of beauty-quark production in pp
  collisions at $ \sqrt{s} $ = 13 TeV via non-prompt D mesons}, JHEP
  \textbf{10}, 110 (2024), \texttt{2402.16417}.
  \doiwoc{10.1007/JHEP10(2024)110}

\bibitem{ALICE-PUBLIC-2025-004}
S.~Acharya et~al. (ALICE), {Preliminary Physics Summary: measurement of
  B$^0$-meson production cross section in proton-proton collisions at $\sqrt{s}
  = 13.6$ TeV} (2025), {ALICE-PUBLIC-2025-004},
  \urlstyle{tt}\url{https://cds.cern.ch/record/2928766}

\bibitem{Cacciari:2012ny}
M.~Cacciari, S.~Frixione, N.~Houdeau, M.L. Mangano, P.~Nason, G.~Ridolfi,
  {Theoretical predictions for charm and bottom production at the LHC}, JHEP
  \textbf{10}, 137 (2012), \texttt{1205.6344}. \doiwoc{10.1007/JHEP10(2012)137}

\bibitem{Benzke:2019usl}
M.~Benzke, B.A. Kniehl, G.~Kramer, I.~Schienbein, H.~Spiesberger, {B-meson
  production in the general-mass variable-flavour-number scheme and LHC data},
  Eur. Phys. J. C \textbf{79}, 814 (2019), \texttt{1907.12456}.
  \doiwoc{10.1140/epjc/s10052-019-7326-y}

\bibitem{Helenius:2023wkn}
I.~Helenius, H.~Paukkunen, {B-meson hadroproduction in the SACOT-m$_{T}$
  scheme}, JHEP \textbf{07}, 054 (2023), \texttt{2303.17864}.
  \doiwoc{10.1007/JHEP07(2023)054}

\bibitem{Barattini:2025wbo}
F.E. Barattini, C.O. Dib, B.~Guiot, {Heavy-hadron production based on
  k$_{t}$-factorization with scale-dependent fragmentation functions}, JHEP
  \textbf{05}, 115 (2025), \texttt{2501.17662}.
  \doiwoc{10.1007/JHEP05(2025)115}

\bibitem{He:2022tod}
M.~He, R.~Rapp, {Bottom Hadrochemistry in High-Energy Hadronic Collisions},
  Phys. Rev. Lett. \textbf{131}, 012301 (2023), \texttt{2209.13419}.
  \doiwoc{10.1103/PhysRevLett.131.012301}

\bibitem{ParticleDataGroup:2024cfk}
S.~Navas et~al. (Particle Data Group), {Review of particle physics}, Phys. Rev.
  D \textbf{110}, 030001 (2024). \doiwoc{10.1103/PhysRevD.110.030001}

\bibitem{ALICE:2023sgl}
S.~Acharya et~al. (ALICE), {Charm production and fragmentation fractions at
  midrapidity in pp collisions at $ \sqrt{s} $ = 13 TeV}, JHEP \textbf{12}, 086
  (2023), \texttt{2308.04877}. \doiwoc{10.1007/JHEP12(2023)086}

\bibitem{ALICE:2021edd}
S.~Acharya et~al. (ALICE), {Prompt and non-prompt J/\ensuremath{\psi}
  production cross sections at midrapidity in proton--proton collisions at
  $\sqrt{s}$ = 5.02 and 13 TeV}, JHEP \textbf{03}, 190 (2022),
  \texttt{2108.02523}. \doiwoc{10.1007/JHEP03(2022)190}

\bibitem{ALICE:2021mgk}
S.~Acharya et~al. (ALICE), {Measurement of beauty and charm production in pp
  collisions at $ \sqrt{s} $ = 5.02 TeV via non-prompt and prompt D mesons},
  JHEP \textbf{05}, 220 (2021), \texttt{2102.13601}.
  \doiwoc{10.1007/JHEP05(2021)220}

\bibitem{ALICE:2023wbx}
S.~Acharya et~al. (ALICE), {Study of flavor dependence of the baryon-to-meson
  ratio in proton--proton collisions at $\sqrt{s}=13$ TeV}, Phys. Rev. D
  \textbf{108}, 112003 (2023), \texttt{2308.04873}.
  \doiwoc{10.1103/PhysRevD.108.112003}

\bibitem{CDF:2004jtw}
D.~Acosta et~al. (CDF), {Measurement of the $J/\psi$ meson and $b-$hadron
  production cross sections in $p\overline{p}$ collisions at $\sqrt{s} = 1960$
  GeV}, Phys. Rev. D \textbf{71}, 032001 (2005), \texttt{hep-ex/0412071}.
  \doiwoc{10.1103/PhysRevD.71.032001}

\bibitem{UA1:1990vvp}
C.~Albajar et~al. (UA1), {Beauty production at the CERN p anti-p collider},
  Phys. Lett. B \textbf{256}, 121 (1991), [Erratum: Phys.Lett.B 262, 497
  (1991)]. \doiwoc{10.1016/0370-2693(91)90228-I}

\bibitem{Catani:2020kkl}
S.~Catani, S.~Devoto, M.~Grazzini, S.~Kallweit, J.~Mazzitelli, {Bottom-quark
  production at hadron colliders: fully differential predictions in NNLO QCD},
  JHEP \textbf{03}, 029 (2021), \texttt{2010.11906}.
  \doiwoc{10.1007/JHEP03(2021)029}

\bibitem{HFLAV:2019otj}
Y.S. Amhis et~al. (HFLAV), {Averages of b-hadron, c-hadron, and $\tau $-lepton
  properties as of 2018}, Eur. Phys. J. C \textbf{81}, 226 (2021),
  \texttt{1909.12524}. \doiwoc{10.1140/epjc/s10052-020-8156-7}

\end{thebibliography}
\end{document}